# Third quantization of the electromagnetic field

J.D. Franson, Physics Dept., University of Maryland Baltimore County, Baltimore, MD 21250 U.S.A.

We consider an approach in which the usual wave function $\psi_j(x_j)$ in the quadrature representation of mode $j$ of the electromagnetic field is further quantized to produce a field operator $\hat{\psi}_j(x_j)$. Since the electromagnetic field is already second quantized, this corresponds to an additional or third quantization. The third-quantization approach can be used to perform certain quantum optics calculations in the Heisenberg picture that could only be performed in the Schrodinger picture when using the conventional second-quantized theory. This approach also allows an interesting generalization of quantum optics and quantum electrodynamics that is analogous to symmetry breaking in elementary particle theory. The predictions of the generalized theory could be tested using a proposed photon scattering experiment.

## I. INTRODUCTION

Particles cannot be created or destroyed in nonrelativistic quantum mechanics. Nevertheless, it is often useful to second-quantize the wave function $\psi(x)$ and its conjugate $\psi^*(x)$ to produce field operators $\hat{\psi}(x)$ and $\hat{\psi}^\dagger(x)$ that can formally annihilate or create particles at position $x$ [1-7]. In this paper, a somewhat analogous approach is introduced in which the usual wave function $\psi_j(x_j)$ [8-12] in the quadrature representation of each mode $j$ of the electromagnetic field is further quantized to produce a field operator $\hat{\psi}_j(x_j)$. This approach allows certain quantum optics calculations to be performed in the Heisenberg picture, in analogy with the use of second-quantized field operators in solid-state physics, for example [13-17].

In quantum optics, each mode of the electromagnetic field is mathematically equivalent to a harmonic oscillator [8,11,18-20]. We can think of each of these harmonic oscillators as containing a single hypothetical particle whose excited states correspond to the presence of photons in the field, as illustrated in Fig. 1. As will be shown below, the operator $\hat{\psi}_j^\dagger(x_j)$ creates additional particles of that kind in the same harmonic oscillator potential, as illustrated in Fig. 2. This generates a hyperspace of the usual Fock space. Since the electromagnetic field is already second quantized, this procedure corresponds to an additional or third quantization [21-26]. For lack of a better term, these hypothetical particles will be referred to as oscillatons [27-29].

The third-quantization approach is equivalent to conventional quantum optics and quantum electrodynamics if we use the standard Hamiltonian, which conserves the number of oscillatons. A generalized theory that is analogous to symmetry breaking in elementary particle theory will also be described. The predictions of the generalized theory could be tested using a proposed photon scattering experiment.

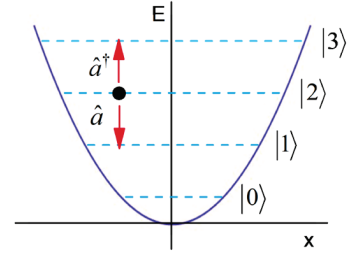

FIG. 1. A harmonic oscillator potential $U(x)$ (blue curve) in one dimension $x$ that contains a single particle represented by a black dot. E is the energy of the particle and the energy eigenstates $|n\rangle$ are represented by dashed lines. The operators $\hat{a}^\dagger$ and $\hat{a}$ increase or decrease the energy of the particle by $\hbar\omega$. A single mode of the electromagnetic field is mathematically equivalent to a harmonic oscillator potential containing a single hypothetical particle whose excited states $|n\rangle$ correspond to $n$ photons in the field [8,11,18-20].

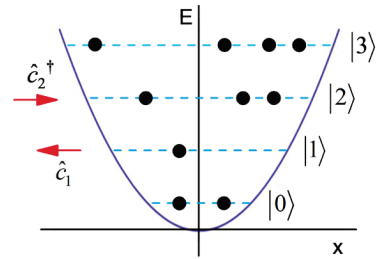

FIG. 2. A harmonic oscillator potential containing $N$ identical bosons represented by black dots. The operators $\hat{c}_n^\dagger$ and $\hat{c}_n$ create or annihilate a particle in the oscillator potential in state $|n\rangle$. The field operator $\hat{\psi}^\dagger(x)$ creates a particle at coordinate $x$. In the case of the electromagnetic field, these hypothetical particles will be referred to as oscillatons.

The remainder of the paper is organized as follows. Section II provides a brief review of the quadrature representation of the electromagnetic field and the usual second-quantization formalism. Those techniques are used in Sec. III to perform an additional or third quantization of the electromagnetic field. Section IV illustrates the use of the third-quantization approach in standard quantum optics by analyzing the decoherence produced by a beam splitter in the Heisenberg picture. An example of a more general theory that does not conserve the number of oscillatons is described in Sec. V, along with a proposed photon scattering experiment that could be used to test the predictions of the theory. A summary and conclusions are provided in Section VI. Additional details of the quantum optics calculations and the generalized theory are given in the appendices.

## II. QUADRATURE REPRESENTATION AND SECOND QUANTIZATION

The second quantization of the normal modes of the classical electromagnetic field results in the usual harmonic oscillator raising and lowering operators $\hat{a}_j^\dagger$ and $\hat{a}_j$ that are responsible for the creation and annihilation of photons [18-19]. The operators $\hat{x}_j$ and $\hat{p}_j$ can then be defined as

$$\begin{aligned}\hat{x}_j &= (\hat{a}_j + \hat{a}_j^\dagger)/\sqrt{2},\\ \hat{p}_j &= -i(\hat{a}_j - \hat{a}_j^\dagger)/\sqrt{2}.\end{aligned} \quad (1)$$

The dimensionless operators $\hat{x}_j$ and $\hat{p}_j$ are referred to as the quadratures of the field in quantum optics [8-10,12], and they are proportional to the electric field of mode $j$ and its time derivative. The quadratures can be directly measured using homodyne techniques [8-12,20,30] and they are used extensively to observe the nonclassical properties of squeezed states [8-11,20,31-32], for example.

For a pure state, the wave function $\psi_j(x_j)$ in the quadrature representation of a single mode $j$ of the second-quantized electromagnetic field can be defined as usual [8-12] by

$$\psi_j(x_j) \equiv \langle x_j | \Psi_j \rangle. \quad (2)$$

Here $|x_j\rangle$ is an eigenstate of $\hat{x}_j$ and $|\Psi_j\rangle$ is the state of mode $j$. For a single-mode field with a definite number $n_j$ of photons (a Fock state), $\psi_j(x_j)$ corresponds to the usual energy eigenfunctions $\phi_n(x_j)$ of a harmonic oscillator that involve the Hermite polynomials [8].

The wave function $\psi_j(x_j)$ gives the probability amplitude that a homodyne measurement will result in that value of the x quadrature. It can be used in the Schrodinger picture to show that postselection based on homodyne measurements can violate Bell's inequality, for example [12,33]. The analysis of experiments of that kind (and others) can be done in the same way in the Heisenberg picture only if the operators $\hat{\psi}_j(x_j)$ and $\hat{\psi}_j^\dagger(x_j)$ are introduced. This may be relevant to an understanding of quantum noise and decoherence in optical amplifiers, for example, which are often analyzed in the Heisenberg picture [34-37].

Before the operators $\hat{\psi}_j(x_j)$ and $\hat{\psi}_j^\dagger(x_j)$ for the electromagnetic field are defined, it may be useful to briefly review the usual second-quantization formalism in nonrelativistic quantum mechanics. We will closely follow the text by Gordon Baym [1]. Consider a harmonic oscillator potential $U(x)$ in one dimension, as illustrated in Fig. 1. The eigenstates $|n\rangle$ of the Hamiltonian are represented by dashed lines. If there is only one particle in the potential, it can be represented by a single black dot occupying one of the eigenstates as in Fig. 1. The usual raising and lowering operators $\hat{a}^\dagger$ and $\hat{a}$ increase or decrease the energy of the particle by one quanta, as indicated by the red arrows.

Even in nonrelativistic quantum mechanics, it is often convenient to introduce an operator $\hat{c}_n^\dagger$ that formally adds or creates an additional particle in eigenstate $|n\rangle$ of the harmonic oscillator potential as illustrated by the red arrow in Fig. 2. Its adjoint $\hat{c}_n$ annihilates a particle if one was there initially. If the particles are bosons, then it is possible to have more than one particle in eigenstate $|n\rangle$. $N_n$ will denote the number of particles in eigenstate $|n\rangle$, and the total number of particles will be denoted by $N$ with no subscript.

It will be assumed that the particles are identical bosons and that they satisfy the commutation relation

$$[\hat{c}_m, \hat{c}_n^\dagger] = \delta_{mn}. \quad (3)$$

The second-quantized field operator $\hat{\psi}(x)$ in the Schrodinger picture is then defined [1] as

$$\hat{\psi}(x) \equiv \sum_n \hat{c}_n \phi_n(x). \quad (4)$$

Here $\phi_n(x)$ is the eigenfunction corresponding to state $|n\rangle$. The fact that the $\phi_n(x)$ form a complete set of orthonormal functions can be combined with Eqs. (3) and (4) to show that

$$[\hat{\psi}(x), \hat{\psi}^{\dagger}(x')] = \delta(x - x'), \quad (5)$$

where $\delta(x - x')$ is the Dirac delta-function. This commutation relation can be used to derive the time dependence of the field operators in the Heisenberg picture.

## III. THIRD QUANTIZATION OF THE FIELD

So far, we have considered the second-quantization of an ordinary harmonic oscillator. The same approach will now be applied to each mode $j$ of the electromagnetic field, which is mathematically equivalent to a harmonic oscillator containing a single hypothetical particle as in Fig. 1 [8,11,18-20]. The excited states $|n_j\rangle$ of the particle correspond to $n_j$ photons in the field, and increasing its energy by $\hbar \omega_j$ and raising the state to $|n_j + 1\rangle$ corresponds to the addition of a photon.

As before, we introduce operators $\hat{c}_{jn}^{\dagger}$ and $\hat{c}_{jn}$ that create or annihilate particles in the harmonic oscillator potential that represents mode $j$ of the field, as in Fig. 2. The operator $\hat{c}_{jn}^{\dagger}$ increases the number of particles in eigenstate $|n_j\rangle$ by one, which increases the dimensions of the usual Fock space to form a hyperspace. The particles will be assumed to be identical bosons, and the commutation relations of Eq. (3) can be used to show that

$$\hat{c}_{jn} |.., N_{jn},.., N_{j0}\rangle = \sqrt{N_{jn}} |..,(N_{jn}-1),..,N_{j0}\rangle,$$
$$\hat{c}_{jn}^{\dagger} |.., N_{jn},.., N_{j0}\rangle = \sqrt{N_{jn}+1} |..,(N_{jn}+1),..,N_{j0}\rangle. \quad (6)$$

Here $|.., N_{jn},.., N_{j0}\rangle$ denotes a state of the oscillator with $N_{jn}$ particles in each of the eigenstates $|n_j\rangle$. This is a generalization of the usual Fock states $|n_j\rangle = |0.., 1_{jn},.., 0_{j0}\rangle$ that correspond to only one particle in the oscillator ($N_j = 1$).

It should be emphasized that these additional particles are not photons. For lack of a better term, we will refer to them as oscillatons. Increasing the energy of an oscillaton by $\hbar \omega_j$ and raising the value of $n$ to $n+1$ corresponds to the addition or emission of a photon. The excited states of the electromagnetic field with $N_j > 1$ will be referred to as hyperphotons. The indices $j$, $n_j$, and $N_{jn}$ correspond to first, second, and third quantization, respectively. The terms oscillaton [27-29], hyperphoton [38-40], and third quantization [21-26] have been used previously with different meanings.

New lowering and raising operators $\hat{a}'_j$ and $\hat{a}'_j{}^{\dagger}$ can be defined as

$$\hat{a}'_j \equiv \sum_{n=1}^{\infty} \sqrt{n}\, \hat{c}_{j,n-1}^{\dagger} \hat{c}_{jn},$$
$$\hat{a}'_j{}^{\dagger} \equiv \sum_{n=0}^{\infty} \sqrt{n+1}\, \hat{c}_{j,n+1}^{\dagger} \hat{c}_{jn}. \quad (7)$$

These operators reduce to the usual raising and lowering operators for $N_j = 1$. The vector potential $\hat{\mathbf{A}}(\mathbf{r})$ can be defined as usual [18] by

$$\hat{\mathbf{A}}(\mathbf{r}) = \sum_{j, \boldsymbol{\varepsilon}_j} \sqrt{\frac{2\pi \hbar c^2}{\omega_j L^3}} \left( \boldsymbol{\varepsilon}_j \hat{a}'_j e^{i \mathbf{k}_j \cdot \mathbf{r}} + \boldsymbol{\varepsilon}_j^* \hat{a}'_j{}^{\dagger} e^{-i \mathbf{k}_j \cdot \mathbf{r}} \right), \quad (8)$$

where $L$ is the length used for periodic boundary conditions, $\boldsymbol{\varepsilon}_j$ are two orthogonal polarization vectors, and $c$ is the speed of light. A similar expression exists for the electric field.

In analogy with Eq. (4), the field operator $\hat{\psi}_j(x_j)$ for mode $j$ of the electromagnetic field can now be defined in the Schrodinger picture as

$$\hat{\psi}_j(x_j) \equiv \sum_n \hat{c}_{jn} \phi_n(x_j). \quad (9)$$

The commutation relation $[\hat{\psi}_j(x_j), \hat{\psi}_j^{\dagger}(x_j')] = \delta(x_j - x_j')$ holds within each mode of the field as in Eq. (5). The field operator $\hat{\psi}_j(x_j)$ could be introduced in a more formal way by using the Lagrangian density and postulating the commutation relation of Eq. (5), but the approach presented here provides more insight.

The use of the terms "second quantization" and "third quantization" may require some clarification. In conventional quantum optics, the first step in the quantization of the electromagnetic field is the determination of the classical normal modes [18]. This has historically been referred to as "first quantization", since it gives rise to a discrete set of frequencies in an optical cavity even though it does not involve quantum mechanics. As a result, the introduction of the photon creation and annihilation operators along with the wave function $\psi_j(x)$ for each mode is generally referred to as "second quantization", even though it is equivalent to the usual first-quantized treatment of a set of independent harmonic oscillators

A similar situation exists in the canonical quantization of the electromagnetic field in the Coulomb gauge. The classical vector potential and electric field are conjugate variables in the classical Lagrangian density. The second-quantization process consists of replacing the classical fields with field operators and then postulating the usual commutation relations. Once again, this process only involves a single quantum-mechanical step even though it is referred to as the second-quantization of the field.

A different situation occurs in the second-quantization of a massive boson in nonrelativistic quantum mechanics. In that case, the introduction of the wave function and Schrodinger's equation is the first quantization step. Second quantization consists of replacing the wave function with field operators and postulating the relevant commutation relations. Thus the second-quantization of a massive boson involves two quantum-mechanical steps, whereas the second-quantization of the electromagnetic field only involves one. This difference is due to the fact that light is already described as a wave or field in classical physics, whereas massive particles are not.

The approach considered here would be equivalent to the usual second-quantization formalism if there were only a single mode of the field. Nevertheless, third quantization appears to be the simplest way to describe the approach, since it involves an additional quantization beyond what is conventionally referred to as the second quantization of the electromagnetic field. In addition, the third-quantization approach is fundamentally different from the usual second-quantization of other fields, such as in the Dirac theory.

## IV. DECOHERENCE IN QUANTUM OPTICS

The use of this approach in conventional quantum optics will now be illustrated by calculating the loss and decoherence produced by a beam splitter. Potential advantages of the approach when applied to more complicated systems are discussed in Appendix A.

Two free-space modes $j$ and $k$ of the electromagnetic field are assumed to be incident on a beam splitter as illustrated in Fig. 3. In the absence of any interaction between the two modes, the Hamiltonian $\hat{H}_0$ can be written in the form

$$\hat{H}_0 = \sum_{n_j=0}^{\infty} \hat{N}_{jn}(n_j+1/2)\hbar\omega_j + \sum_{n_k=0}^{\infty} \hat{N}_{kn}(n_k+1/2)\hbar\omega_k. \quad (10)$$

Here $\hat{N}_{jn} = \hat{c}_{jn}^{\dagger}\hat{c}_{jn}$ and $\hat{N}_{kn} = \hat{c}_{kn}^{\dagger}\hat{c}_{kn}$.

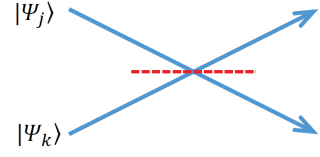

FIG. 3. Two beams of light incident on a beam splitter.

The effects of the beam splitter coupling can be described by an interaction potential $U'(x_j, x_k) = \varepsilon x_j x_k$ where $\varepsilon$ is a real constant. Using the definition of the quadratures in Eq. (1), it can be seen that $U'(x_j, x_k)$ will involve $\hat{a}_j^{\dagger}\hat{a}_k$ and $\hat{a}_k^{\dagger}\hat{a}_j$, which can transfer photons from one mode to the other. The interaction Hamiltonian $\hat{H}'$ can be written in terms of the field operators as [1]

$$\hat{H}' = \iint dx_j dx_k \hat{\psi}_j^{\dagger}(x_j)\hat{\psi}_k^{\dagger}(x_k) \\ \times U'(x_j, x_k)\hat{\psi}_k(x_k)\hat{\psi}_j(x_j). \quad (11)$$

A continuous coupling of this kind occurs in two nearby wave guides due to their evanescent fields, for example, which is equivalent to a beam splitter.

In the Heisenberg picture, the time dependence of the operator $\hat{c}_{jn}(t)$ is given by

$$\frac{d\hat{c}_{jn}(t)}{dt} = \frac{1}{i\hbar}\left[\hat{c}_{jn}(t), \hat{H}\right], \quad (12)$$

with $\hat{H} = \hat{H}_0 + \hat{H}'$. The commutator in Eq. (12) can be evaluated using Eqs. (3), (9), (10), and (11) combined with the identity

$$\int_{-\infty}^{\infty} \phi_{n'}^{*}(x)x\phi_n(x)dx = \frac{1}{\sqrt{2}}\left(\sqrt{n}\delta_{n',n-1} + \sqrt{n+1}\delta_{n',n+1}\right). \quad (13)$$

This gives

$$i\hbar\frac{d\hat{c}_{jn}(t)}{dt} = \left(n_j + \frac{1}{2}\right)\hbar\omega_j \hat{c}_{jn} \\ + \frac{\varepsilon}{2}\sqrt{n}\,\hat{c}_{j,n-1}\sum_{m=1}^{\infty}\sqrt{m}\,\hat{c}_{k,m-1}^{\dagger}\hat{c}_{km} \quad (14) \\ + \frac{\varepsilon}{2}\sqrt{n+1}\,\hat{c}_{j,n+1}\sum_{m=0}^{\infty}\sqrt{m+1}\,\hat{c}_{k,m+1}^{\dagger}\hat{c}_{km}.$$

Two additional terms in Eq. (14) that do not conserve energy have been neglected in the usual rotating-wave approximation [8,20]. The time rate of change of $\hat{c}_{kn}(t)$ is given by a similar expression. This set of coupled equations can be solved to find the form of these operators and their adjoints as a function of time in the Heisenberg picture. The results can then be inserted into Eq. (9) to obtain the form of the field operator.

Eq. (14) and the corresponding equation for $\hat{c}_{kn}(t)$ were solved numerically using the hyperphoton number states $|..,N_{jn},...,N_{j0}\rangle \otimes |...N_{kn},...,N_{k0}\rangle$ as a basis for a matrix representation. Given the order of the operators in Eq. (14), it is only necessary to include the states with $N_j$ and $N_k$ equal to 0 or 1 (the initial state corresponds to $N_j = N_k = 1$), as is discussed in more detail in Appendix B. We will denote the state with no oscillatons in mode $j$ by $|Z_j\rangle = |..,0_{jn},..,0_{j0}\rangle$, with a similar expression for mode $k$. The relevant Hilbert subspace for this example includes $|Z_j\rangle$ and $|Z_k\rangle$ in addition to the usual Fock states, and it is only slightly larger than the Hilbert space for conventional quantum optics. The numerical calculations are described in more detail in Appendix B.

Fig. 4(a) shows the calculated probability density $P(x_j,t) = \langle \hat{\psi}_j^\dagger(x_j,t)\hat{\psi}_j(x_j,t)\rangle$ as a function of time for the case in which the initial state in mode $j$ was a coherent state [8-11] with a mean photon number of $\bar{n}_j = 4$. Mode $k$ was assumed to initially be in its vacuum state $|0_k\rangle$ with no photons. It can be seen that the probability density in mode $j$ is described by a Gaussian distribution whose mean displacement oscillates sinusoidally with a decreasing amplitude. Fig. 4(b) shows the corresponding results for mode $k$, whose amplitude increases at the expense of mode $j$ due to the beam splitter coupling [41].

One of the advantages of using the third-quantized Heisenberg picture is that $\hat{\psi}_j(t)$ and $\hat{\psi}_k(t)$ only need to be calculated once, after which they can be used with any initial state or measurement. A more interesting example corresponds to the case in which the initial state in mode $j$ is assumed to be a Schrodinger cat state given by

$$|\Psi_j\rangle = c_n \left(|\alpha\rangle + e^{i\theta}|-\alpha\rangle\right). \quad (15)$$

Here $c_n \approx 1/\sqrt{2}$ is a normalization constant, $|\alpha\rangle$ is a coherent state with a real amplitude $\alpha$, and $\theta$ is an arbitrary phase shift. Mode $k$ was assumed to initially be in its vacuum state once again. The joint probability density $P_J(x_j,x_k,t)$ can be calculated using

$$P_J(x_j,x_k,t) = \langle \hat{\psi}_j^\dagger(x_j,t)\hat{\psi}_k^\dagger(x_k,t)\hat{\psi}_k(x_k,t)\hat{\psi}_j(x_j,t)\rangle. \quad (16)$$

This is plotted in Fig. 5(a) at the initial time and then again at a later time in Fig. 5(b) after the two beams have passed through the beam splitter. The entanglement produced by the beam splitter [12] can be seen from the fact that $x_j$ and $x_k$ become correlated.

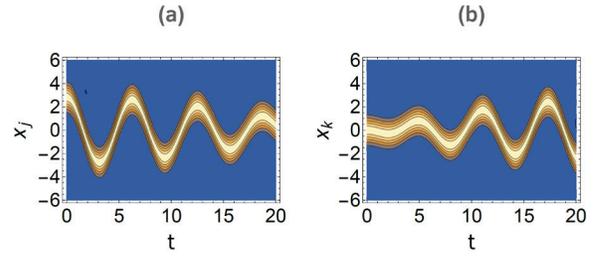

FIG. 4. Effects of a beam splitter on a coherent state incident in mode $j$ with mode $k$ initially in its vacuum state. (a) Probability density $P(x_j,t) = \langle \hat{\psi}_j^\dagger(x_j,t)\hat{\psi}_j(x_j,t)\rangle$ for mode $j$ plotted as a function of the quadrature $x_j$ and the time $t$. (b) Probability density $P(x_k,t)$ for mode $k$. The width of these probability distributions is due to vacuum fluctuations. (Arbitrary units.)

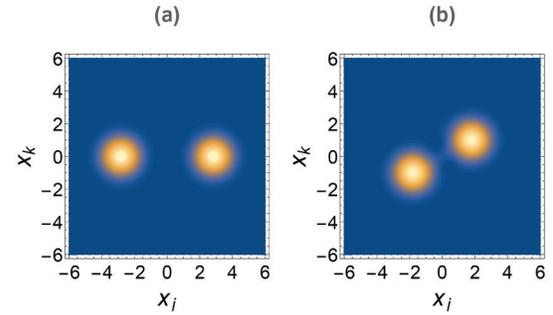

FIG. 5. Joint probability density $P_J(x_j,x_k,t)$ for a Schrodinger cat state plotted as a function of the quadratures $x_j$ and $x_k$. (a) Joint probability distribution evaluated before the beam splitter ($t=0$). (b) Entangled state produced by the beam splitter [12] with $\varepsilon = 0.12$ and $t = 12$. (Arbitrary units.)

There are proposed experiments in quantum optics [12,33,37,42] that could measure the expectation value of a coherence operator $\hat{C}_j(x_j,\Delta)$ defined in the Heisenberg picture by

$$\hat{C}_j(x_j,\Delta) \equiv \frac{1}{2}\left[\hat{\psi}_j^\dagger(x_j+\Delta)\hat{\psi}_j(x_j-\Delta)\right] + h.c. \quad (17)$$

The proposed experiments apply a displacement in phase space in one arm of an interferometer. This causes the observable probability density $\psi^*\psi$ in the Schrodinger picture to contain interference cross terms terms such as $\psi_0^*(x_1)\psi_0(x_2)$ [37], where $\psi_0(x)$ is the initial wave function while $x_1$ and $x_2$ are two different points in phase space. Interference of this kind can be measured using homodyne techniques, and it can be analyzed in the Schrodinger picture using the wave function itself or quasiprobability distributions based on the wave function [37, 42].

Since the Heisenberg picture is based on the use of operators, interference of this kind between two different points in phase space can be analyzed in the Heisenberg picture only if the wave function $\psi(x)$ is replaced with an operator $\hat{\psi}(x)$. The product of the two field operators in Eq. (17) is equivalent to the interference term $\psi_0^*(x_1)\psi_0(x_2)$ in the Schrodinger picture. This requires the third-quantization approach described above, and the description of quantum interference of this kind in the Heisenberg picture was the original motivation for this paper.

With a suitable choice of the parameter $\Delta$, a measurement of the expectation value of $\hat{C}_j(0,\Delta)$ can determine the amount of potential quantum interference (coherence) between the two components of the Schrodinger cat state of Eq. (15). This is illustrated in Fig. 6, where $\langle\hat{C}_j(0,\Delta)\rangle$ in the third-quantized Heisenberg picture is plotted as a function of the phase difference $\theta$. The calculations were performed numerically, as described in Appendix B. The blue (solid) curve shows the interference pattern as measured before the beam splitter, while the red (dashed) curve shows the corresponding results after a beam splitter with the same parameters as in Fig. 5(b). The value of $\Delta$ was chosen to maximize the amount of interference in both cases.

Operator $\hat{C}_j(x_j,\Delta)$ measures the coherence of the electromagnetic field between two different points in quadrature space, while earlier coherence functions measure it between different points in space-time [8,20]. As discussed above, the decoherence shown in Fig. 6 cannot be calculated in the Heisenberg picture without using the third-quantization approach. Roughly speaking, elementary quantum mechanics requires the wave function $\psi(x)$ as well as the operator $\hat{x}$ to calculate the results of experiments, and a complete description of quantum optics in the Heisenberg picture requires the operator $\hat{\psi}(x)$ in addition to $\hat{x}$.

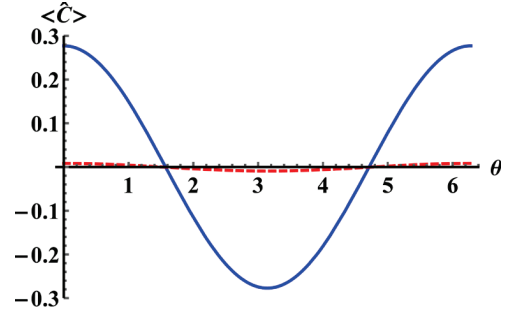

FIG. 6. Quantum interference between the two components of a Schrodinger cat state. The blue (solid) curve shows the expectation value of operator $\hat{C}_j(0,\Delta)$ as a function of $\theta$ evaluated before the beam splitter. The red (dashed) curve shows the expectation value of $\hat{C}_j(0,\Delta)$ after the beam splitter using the same parameters as in Fig. 5(b). It can be seen that passing a Schrodinger cat state through a beam splitter will reduce the quantum interference by an amount that is much larger than the reduction in the field amplitude [42]. (Arbitrary units.)

A similar situation exists for optical parametric amplifiers, which are often analyzed in the Heisenberg picture [36]. As a result, the usual linear relationship between the input and output quadrature operators in the Heisenberg picture cannot describe the decoherence produced by a parametric amplifier, as is discussed in more detail in Appendix A.

The third-quantization approach may also be useful when a full multi-mode analysis of interacting optical pulses with a continuous range of frequencies is required, since the field operators remain a function of only one coordinate while the corresponding quasiprobability distributions would be a function of an infinite number of coordinates. This situation is somewhat similar to the use of field operators in solid-state physics to avoid wave functions that depend on a large number of electron coordinates, as is discussed in more detail in Appendix A. Further investigations will be required to determine the practical value of this approach in calculations of that kind.

## V. GENERALIZED THEORY

The third-quantization approach allows an interesting generalization of quantum optics and quantum electrodynamics in which the number of oscillatons is not conserved. Here we consider one example of a generalized theory of that kind, and we propose a photon scattering experiment that could be used to set an upper bound on the effects that it predicts.

The interaction Hamiltonian of Eq. (11) conserves the number of oscillatons and agrees with conventional quantum optics, but in principle there could be other Hamiltonians that do not, such as

$$\hat{H}' = -\frac{1}{c}\int d^3\mathbf{r}\,\hat{\mathbf{j}}(\mathbf{r})\cdot\hat{\mathbf{A}}'(\mathbf{r}). \qquad (18)$$

Here $\hat{\mathbf{j}}(\mathbf{r})$ is the second-quantized current associated with another particle, such as an electron, while $\hat{\mathbf{A}}'(\mathbf{r})$ is the vector potential defined in Eqs. (7) and (8) with $\hat{c}_{jn}$ and $\hat{c}_{jn}^\dagger$ replaced using a Bogoliubov transformation [14-17,43-47] given by

$$\begin{aligned}\hat{c}_{jn} \to \hat{c}'_{jn} &= \beta\left(\cos\gamma\,\hat{c}_{jn} + \sin\gamma\,\hat{c}_{jn}^\dagger\right) \\ \hat{c}_{jn}^\dagger \to \hat{c}'^{\dagger}_{jn} &= \beta\left(\sin\gamma\,\hat{c}_{jn} + \cos\gamma\,\hat{c}_{jn}^\dagger\right).\end{aligned} \qquad (19)$$

Eq. (18) can also be written in a covariant form in the Lorentz gauge [18].

The constant $\beta = 1/(\cos^2\gamma - \sin^2\gamma)^{1/2}$ maintains the commutation relations while $\gamma$ is an unknown angle similar to the mixing angles that occur in elementary particle theory [48-50]. Bogoliubov transformations commonly occur in quantum optics [46], superconductivity [16,17], and general relativity [47], and Eq. (19) appears to be the simplest generalization of quantum optics and quantum electrodynamics based on the third-quantization approach.

Although Eq. (19) is intended to be an arbitrary example of a generalized theory, it can be derived under the assumption that the oscillatons interact with another hypothetical boson B with a large mass $M$. If $\hat{b}_j$ and $\hat{b}_j^\dagger$ are the annihilation and creation operators for particle B in mode $j$, then we can consider an interaction Hamiltonian of the form

$$\hat{H}_B' = \varepsilon\sum_{jn}\left(\hat{b}_{jn} + \hat{b}_{jn}^\dagger\right)\left(\hat{c}_{jn} + \hat{c}_{jn}^\dagger\right), \qquad (20)$$

where $\varepsilon \ll 1$ is an unknown constant. This interaction Hamiltonian has the same form as a coupling between the displacement of two harmonic oscillators. The Bogoliubov transformation of Eq. (19) can be derived from Eq. (20) in the limit of $Mc^2 \gg \hbar\omega$ and $\varepsilon \ll 1$, as is shown in Appendix C.

The interaction Hamiltonian of Eq. (20) breaks the symmetry that would otherwise conserve the number of oscillatons. A discussion of the possible nature of these particles along with their connection to symmetry breaking in elementary particle physics [50] can be found in Appendix C. Although this model is speculative, it does provide some motivation for the assumed form of the Bogoliubov transformation of Eq. (19).

The generalized theory of Eqs. (18) and (19) can be tested experimentally using the photon scattering experiment shown in Fig. 7. Inserting the Bogoliubov transformation of Eq. (19) into the vector potential operator in Eqs. (8) or (18) will produce terms that involve two oscillaton creation operators or two oscillaton annihilation operators. As a result, $\hat{H}'$ can create or annihilate a pair of oscillatons along with the emission or absorption of a photon if $\gamma \neq 0$, while it reduces to the standard interaction Hamiltonian in the Coulomb gauge [18] for $\gamma = 0$.

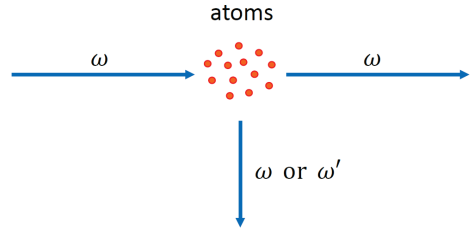

FIG. 7. Proposed photon scattering experiment to test the generalized theory of Eq. (19). Photons at frequency $\omega$ are incident on a cloud of two-level atoms. Some of the photons are scattered through a 90° angle with final frequencies of $\omega$ or $\omega' = \omega/2$. Energy is conserved in the latter case by the creation of a pair of oscillatons. The predicted ratio of the two scattering rates is given by Eq. (21), which can be used to set an upper bound on the mixing angle $\gamma$.

In the process shown in Fig. 7, an incident photon of frequency $\omega$ is absorbed into a virtual state in the usual way, after which a single scattered photon with frequency $\omega'$ is emitted along with the creation of a pair of oscillatons at the new frequency. The atoms are left in their original (ground) state. As shown in Appendix D, energy conservation requires that $\omega' = \omega/2$, which corresponds to a subharmonic or inelastic scattering process.

We can define the ratio R as the subharmonic scattering rate at frequency $\omega'$ divided by the usual elastic scattering rate at frequency $\omega$. As shown in Appendix D, this ratio is predicted by the theory to be given by the simple expression

$$R = 4\gamma^2 \qquad (21)$$

for $|\gamma| \ll 1$. The two scattering rates can be measured using appropriate filters, and an experiment of this kind could set an upper bound on the value of $\gamma$.

It has been tacitly assumed that the oscillaton mass is zero since a photon has zero mass, but an experiment of this kind could also determine the mass of the oscillaton as discussed in Appendix D. High-energy experiments involving particle accelerators or cosmic rays would be required if the mass of the oscillaton is very large.

Other experimental tests of the generalized theory may also be possible, since the existence of oscillatons would be expected to increase the decay rate of excited atoms or more exotic systems such as muonium. Oscillatons could conceivably play a role in the discrepancy observed in recent measurements of the fine structure of positronium [51,52], for example, since Eqs. (18) and (19) would contribute additional Feynman diagrams. These topics require further investigation and are beyond the intended scope of this paper.

## VI. SUMMARY AND CONCLUSIONS

A third-quantization approach has been introduced in which the usual wave function $\psi_j(x_j)$ for each mode $j$ of the second-quantized electromagnetic field is further quantized to produce a field operator $\hat{\psi}_j(x_j)$. The operator $\hat{\psi}_j^\dagger(x_j)$ creates an additional hypothetical particle (oscillaton) in the harmonic oscillator corresponding to mode $j$ of the electromagnetic field, where the emission or absorption of a photon corresponds to a change in the energy level of an oscillaton as illustrated in Fig. 2. The theory is equivalent to conventional quantum optics and quantum electrodynamics if we use the standard Hamiltonian, which conserves the number of oscillatons.

The third-quantization approach can be used to perform certain quantum optics calculations in the Heisenberg picture that could only be performed in the Schrodinger picture when using the conventional second-quantized theory. As a result, a complete description of quantum optics in the Heisenberg picture requires the use of the third-quantization approach. This can provide additional insight into systems that have often been analyzed in the Heisenberg picture, such as the input-output relations for an optical parametric amplifier. Other phenomena that can be analyzed in this way include coherence operators, quantum interference, and postselection in quadrature space. The third quantization approach may also be useful when analyzing a continuum of modes, where the usual quasiprobability distributions would depend on an infinite number of coordinates.

A specific example of a generalized theory that does not conserve the number of oscillatons was also described. The theory is based on a Bogoliubov transformation that couples the oscillaton creation and annihilation operators through an unknown mixing angle $\gamma$. This form of the Bogoliubov transformation can be derived from the assumption that the oscillatons interact with a hypothetical boson with a large mass, and the theory is analogous to symmetry breaking in elementary particle theory. A photon scattering experiment was proposed that could set an upper bound on the value of the mixing angle and determine the mass of the oscillaton.

In summary, the third quantization of the electromagnetic field may be a useful alternative for certain calculations in quantum optics, while allowing an interesting generalization of quantum optics and quantum electrodynamics that could be tested experimentally.

## ACKNOWLEDGEMENTS

This work was supported in part by the National Science Foundation under grant number PHY-1802472.

## APPENDIX A: APPLICATIONS IN QUANTUM OPTICS

The third-quantization approach allows certain calculations in quantum optics to be performed in the Heisenberg picture, while the corresponding calculations could only be performed in the Schrodinger picture when using the conventional second-quantization approach. One might ask whether or not the third-quantization approach is of any practical use, given that the same results could be obtained using the Schrodinger picture. In this appendix, we argue that the third-quantization approach can provide additional insight into certain phenomena that have been traditionally analyzed in the Heisenberg picture, giving results that were incomplete or potentially misleading. We also compare the third-quantization approach with the use of quasiprobability distributions, and argue that third quantization may have some potential advantages when analyzing systems with a continuum of modes.

As discussed in the text, the decoherence of a Schrodinger cat state passing through a beam splitter, as plotted in Fig. 6, cannot be calculated in the usual second-quantized Heisenberg picture [42]. This situation can be understood intuitively from the fact that, in the Schrodinger picture, the wave function $\psi(x)$ gives the probability amplitude that a homodyne measurement will result in that particular value of the quadrature. What gives the probability amplitude of obtaining a particular value of x from a homodyne measurement in the Heisenberg picture? Only the third-quantized field operator $\hat{\psi}(x,t)$ can do that, not the operator $\hat{x}(t)$. As a result, there are experiments in quantum optics that cannot be analyzed in the usual second-quantized Heisenberg picture. This limitation on the use of the conventional Heisenberg picture [37,42] does not appear to be widely appreciated.

An important example of this is the decoherence produced by a linear optical amplifier, which is commonly analyzed in the Heisenberg picture based on the pioneering work by Caves and others [34-36]. As is well known, the input and output quadrature operators are related by a simple transformation given by

$$\hat{x}_{out} = g\hat{x}_{in} + \hat{N}_{noise}. \tag{A1}$$

Here $\hat{x}_{in}(t)$ and $\hat{x}_{out}(t)$ are the input and output quadratures in the usual Heisenberg picture, $g$ is the gain of the amplifier, and $\hat{N}_{noise}$ is a quantum noise operator. There are situations where $g \to 1$ and $\hat{N}_{noise} \to 0$ even though there is an exponential decrease in the coherence of a cat state [37]. Eq. (A1) would seem to imply that the output field is the same as the input in that case, despite the large decoherence.

This example suggests that the third-quantized field operators $\hat{\psi}_j(x_j,t)$ and $\hat{C}_j(x_j,\Delta,t)$ provide a more complete description of the system than $\hat{x}_j(t)$ does alone, especially for entangled states. Once again, the reason is that the probability amplitude for obtaining a specific value of $x$ from a homodyne measurement can be found from $\hat{\psi}_j(x_j,t)$ but not from the usual operator $\hat{x}_j(t)$. The decoherence produced by an optical parametric amplifier can be calculated in the Heisenberg picture using the third-quantization approach in the same way that the decoherence of a beam splitter was calculated in the text. The corresponding results cannot be obtained using the familiar linear transformation of Eq. (A1).

Many problems in quantum optics can be solved using quasiprobability distributions, such as the Wigner distribution [41]. The displacement by $\pm\Delta$ in Eq. (17) for the operator $\hat{C}_j(x_j,\Delta)$ is similar in appearance to the displacement of the wave function by $\pm y/2$ in the Wigner distribution, which is defined [41] by

$$W(x_j, p_j) \equiv \frac{1}{2\pi} \int_{-\infty}^{\infty} dy\, e^{-ip_j y} \\ \times \psi_j^*\left(x_j - \frac{1}{2}y\right)\psi_j\left(x_j + \frac{1}{2}y\right). \tag{A2}$$

Here we are considering a pure state of a single mode $j$ of the electromagnetic field. But the definition of operator $\hat{C}_j(x_j,\Delta)$ does not include the exponential factor involving $p_j$ or the integral that appears in the definition of the Wigner distribution. As a result, $\hat{C}_j(x_j,\Delta)$ is defined in quadrature space rather than phase space. $\hat{C}_j(x_j,\Delta)$ is more closely related to a coherence function than a quasiprobability distribution. It could be normalized in the usual way and higher-order coherence functions can be defined for $N_j > 1$.

A more significant difference between $\hat{C}_j(x_j,\Delta,t)$ and the Wigner distribution can be seen if there are two or more modes that interact, as in the beam splitter example in the text. In that case the third-quantized field operator $\hat{\psi}_j(x_j,t)$ and operator $\hat{C}_j(x_j,\Delta,t)$ include all of the effects of the entanglement with the other mode $k$, even though they are only a function of one coordinate. In contrast, the wave function $\psi(x_j, x_k, t)$ would depend on both coordinates and the two-mode Wigner distribution would be a function of $\hat{x}_j$, $\hat{x}_k$, $\hat{p}_j$, and $\hat{p}_k$. Eqs. (17) and (A2) would have a very different form in that case.

The use of third-quantized field operators may have some advantages compared to using quasiprobability distributions if we need to do a full multi-mode analysis of interacting optical pulses with a continuous range of frequencies [46]. In that case, there would be an infinite number of interacting modes and the entangled wave function and quasiprobability distributions would all be a function of an infinite number of coordinates. In contrast, each of the field operators $\hat{\psi}_j(x_j,t)$ would still be a function of only one coordinate. This situation is somewhat analogous to the use of field operators in solid-state physics, where the density of electrons can be described by a field operator that is a function of only one coordinate, rather than a wave function that depends on a very large number of electron coordinates. The use of the third-quantization approach for problems of this kind appears to be promising but it requires further investigation.

## APPENDIX B: NUMERICAL CALCULATIONS

Analytic solutions to Eq. (14) would be desirable, but our earlier work [12] on systems of this kind using the Schrodinger picture did not allow analytic solutions in general and that is probably the case for the third-quantization approach as well. Perturbation theory could be used in many other applications of interest, such as quantum electrodynamics, but the interaction is not small in the situation of interest here and perturbation theory cannot be used. As a result, Eq. (14) for $\dot{\hat{c}}_{jn}(t)$ along with the corresponding equation for $\dot{\hat{c}}_{kn}(t)$ were solved numerically instead.

The hyperphoton number states $|..,N_{jn},..,N_{j0}\rangle \otimes |..,N_{kn},..,N_{k0}\rangle$ were used as a basis for a matrix representation of the operators of interest. As discussed in the text, the only relevant states in the examples of interest here correspond to $N_j$ and $N_k$ equal to 0 or 1, with both values equal to 1 in the initial state. The state with zero oscillatons is required because the field operator $\hat{\psi}_j(x_j - \Delta)$ in the definition of the coherence operator of Eq. (17) can act on a physical state with one oscillaton to temporarily create a state with no oscillatons, after which $\hat{\psi}_j^\dagger(x_j + \Delta)$ recreates the oscillaton at another location. This kind of situation frequently occurs in other applications of the second-quantization formalism as well, such as solid-state physics.

The mean number of photons in the initial coherent state of mode $j$ was chosen to be $\bar{n}_j = 4$, with the other mode initially in the vacuum state. Since the probability amplitude drops off exponentially with increasing photon number in a coherent state, it was sufficient to cut off the state vector at a maximum number of photons equal to $n_{max} = 16$ in both modes. The value of $n_{max}$ was varied to ensure that the cutoff had no significant effect on the results.

With the addition of the state $|Z_j\rangle$ and the usual vacuum state $|0_j\rangle$, the total number of elements in the state vector for mode $j$ alone was equal to $n_{max} + 2 = 18$. With the same number of elements in mode $k$, the dimensions of the combined Hilbert space was $(n_{max} + 2)^2 = 324$. The number of elements in the matrix representation of each of the operators was the square of that, or 104,976. All of the matrices were very sparse and the memory requirements as well as the execution time were greatly reduced using Mathematica's sparse matrix routines.

In order to put the operators in the form of a matrix, it was useful to label each element of the combined state vector with a single index $l$ that ranged from 1 to 324. The way in which the states are labeled is arbitrary, but a suitable choice for the labelling allowed the number of photons in each mode to be written as a simple function of $l$, for example. That in turn allowed the nonzero values of $\hat{c}_{jn}[l',l,t]$ and $\hat{c}_{kn}[l',l,t]$ to be specified at the initial time $t = t_0$ in a straightforward way using Eq. (6).

The matrices $\hat{c}_{jn}[l',l,t]$ and $\hat{c}_{kn}[l',l,t]$ were then incremented over small time intervals $\Delta t$ using the fourth-order Runge-Kutta algorithm with derivatives given by Eq. (14). Since the residual errors in the Runge-Kutta algorithm are on the order of $\Delta t^5$, the results converged rapidly and had no significant dependence on the choice of the time step. The results shown in the text were based on the use of 1200 time steps. Built-in Mathematica routines such as NDSolve were not used because they store the results at all of the intermediate steps and require a large amount of memory as a result.

Once $\hat{c}_{jn}[l',l,t]$ and $\hat{c}_{kn}[l',l,t]$ had been calculated, they were inserted into Eq. (9) for the field operator which was then used to calculate the expectation values of interest (using the initial state). The calculations shown in the text required approximately 10 min of computer time and 250 MB of memory on a personal computer.

## APPENDIX C: BOGOLIUBOV TRANSFORMATION

The Bogoliubov transformation of Eq. (19) will be derived in this Appendix starting from the interaction Hamiltonian of Eq. (20). The possible nature of these particles will also be discussed, including an analogy with the symmetry breaking responsible for neutrino oscillations.

The Bogoliubov transformation of Eq. (19) was motivated by phenomena in quantum optics [46], superconductivity [16,17], and general relativity [47]. A Bogoliubov transformation can often be used to diagonalize an interaction Hamiltonian, and Eq. (19) can be viewed as the result of an interaction between the oscillatons and another hypothetical particle B as discussed in the text. The bare oscillatons would no longer correspond to the true eigenstates of the system as a result of the interaction. The angle $\gamma$ is somewhat analogous to a mixing angle, such as the Cabibbo angle [48] or the Weinberg angle [49].

In the interaction picture, the state $|\psi(t)\rangle$ of the system at time $t$ can be related to the initial state

$|\psi(0)\rangle$ using time-dependent perturbation theory, which gives

$$\begin{aligned}|\psi(t)\rangle &= \hat{U}(t)|\psi(0)\rangle \\ &= \left[\hat{U}^{(0)} + \hat{U}^{(1)} + \hat{U}^{(2)} + ...\right]|\psi(0)\rangle.\end{aligned} \quad \text{(C1)}$$

Here the transition matrices $\hat{U}^{(0)}$, $\hat{U}^{(1)}$, and $\hat{U}^{(2)}$ are given by

$$\begin{aligned}\hat{U}^{(0)} &= \hat{I}, \\ \hat{U}^{(1)} &= \frac{1}{i\hbar}\int_{-\infty}^{t}dt'\hat{H}_B'(t'), \\ \hat{U}^{(2)} &= \frac{1}{(i\hbar)^2}\int_{-\infty}^{t}dt'\int_{-\infty}^{t'}dt''\hat{H}_B'(t')\hat{H}_B'(t'').\end{aligned} \quad \text{(C2)}$$

The initial time has been taken to be $-\infty$ and a factor of $\exp[\eta t]$ will be included in $\hat{H}_B'$ with the limit $\eta \to 0$ taken at the end of the calculation as usual [1] to ensure a slow turn-on of the interaction.

Operators in the interaction picture will be labelled with a subscript I, while those in the Schrodinger picture will be labelled with a subscript S. It will be convenient to consider a specific value of the coefficients $j$ and $n$, and to drop those subscripts in what follows. Particle B will be assumed to initially be in its vacuum state, so that the first-order contribution reduces to

$$\hat{U}^{(1)} = \frac{\varepsilon}{i\hbar}\int_{-\infty}^{t}dt'e^{\eta t'}\hat{b}_I^{\dagger}(t')\left[\hat{c}_I(t') + \hat{c}_I^{\dagger}(t')\right]. \quad \text{(C3)}$$

The second-order contribution can then be written as

$$\begin{aligned}\hat{U}^{(2)} &= \frac{\varepsilon^2}{(i\hbar)^2}\int_{-\infty}^{t}dt'e^{\eta t'}\hat{b}_I(t')\left[\hat{c}_I(t') + \hat{c}_I^{\dagger}(t')\right] \\ &\times \int_{-\infty}^{t'}dt''e^{\eta t''}\hat{b}_I^{\dagger}(t'')\left[\hat{c}_I(t'') + \hat{c}_I^{\dagger}(t'')\right].\end{aligned} \quad \text{(C4)}$$

Second-order terms involving $(b_I^{\dagger})^2$ have been omitted since they correspond to a virtual state containing two B particles, which is negligible in the limit of large $M$.

The relevant operators in the interaction picture are related to those in the Schrodinger picture by

$$\begin{aligned}\hat{c}_I(t) &= e^{-i\omega t}\hat{c}_S, \\ \hat{b}_I(t) &= e^{-i\Omega t}\hat{b}_S,\end{aligned} \quad \text{(C5)}$$

with similar expressions for the adjoint operators. Here the angular frequency $\Omega$ has been defined as $\hbar\Omega = Mc^2$. The second line of Eq. (C5) is valid when $Mc^2 \gg \hbar\omega$ so that the rest mass of particle B is larger than its kinetic energy in the oscillator level $n$.

Inserting Eq. (C5) into Eqs. (C3) and (C4) and performing the integrals gives

$$\begin{aligned}\hat{U}^{(1)} &= -\frac{\varepsilon}{\hbar\Omega}\hat{b}_I^{\dagger}\left[\hat{c}_I + \hat{c}_I^{\dagger}\right] \\ \hat{U}^{(2)} &= -\frac{\varepsilon^2}{2(\hbar\Omega)(\hbar\omega)}\left[\hat{c}_I^2 - \hat{c}_I^{\dagger 2}\right].\end{aligned} \quad \text{(C6)}$$

Terms involving $\hat{c}_I^{\dagger}\hat{c}_I$ have been dropped, since they correspond to phase shifts due to a small shift in the perturbed energy, which is of no interest here.

The same expectation values (observables) would be obtained if we use the initial value $|\psi(0)\rangle$ of the state vector and introduce new operators given by

$$\begin{aligned}\hat{c}' &= \hat{U}^{\dagger}\hat{c}_I\hat{U}, \\ \hat{c}'^{\dagger} &= \hat{U}\hat{c}_I^{\dagger}\hat{U}^{\dagger}.\end{aligned} \quad \text{(C7)}$$

Using the identity $\hat{A}\hat{B} = \hat{B}\hat{A} + \left[\hat{A}, \hat{B}\right]$ for any two operators $\hat{A}$ and $\hat{B}$, along with the fact that $\hat{U}$ is unitary, allows Eq (C7) to be rewritten as

$$\hat{c}' = \hat{c}_I + \left[\hat{U}^{\dagger}, \hat{c}_I\right]\hat{U} \quad \text{(C8)}$$

Evaluating the commutator using Eq. (C6) gives

$$\hat{c}' = \hat{c}_I + \frac{\varepsilon^2}{(\hbar\omega)(\hbar\Omega)}\hat{c}_I^{\dagger}. \quad \text{(C9)}$$

to second order. The first-order correction and additional terms in the second-order correction can be shown to be negligible for $\hbar\Omega \gg \hbar\omega$.

Eq. (C9) shows that the theory predicts a coupling between $\hat{c}$ and $\hat{c}^{\dagger}$. For $\gamma \ll 1$, $\cos\gamma \approx 1$, $\sin\gamma \approx \gamma$, and $\beta \approx 1$. In that case, Eq. (C9) is equivalent to the Bogoliubov transformation of Eq. (19) with $\gamma$ given by

$$\gamma = \frac{\varepsilon^2}{(\hbar\omega)(\hbar\Omega)}. \quad \text{(C10)}$$

This result shows that the Bogoliubov transformation that mixes the oscillaton creation and annihilation operators can be viewed as the result of an interaction of the oscillatons with a massive boson B.

The form of the interaction Hamiltonian in Eq. (20) suggests that the oscillaton and B particles must have similar properties, and that they may be members of a family of particles with three different generations, as are leptons and quarks. In that case, there would be a third boson B' with mass $M'$ in the same family. Eq. (20) could then be generalized to

$$\hat{H}_B' = \varepsilon_1 \sum_{jn} \left(\hat{b}_{jn} + \hat{b}_{jn}^\dagger\right)\left(\hat{c}_{jn} + \hat{c}_{jn}^\dagger\right)$$
$$+\varepsilon_2 \sum_{jn} \left(\hat{b}'_{jn} + \hat{b}'_{jn}^\dagger\right)\left(\hat{c}_{jn} + \hat{c}_{jn}^\dagger\right) \quad \text{(C11)}$$
$$+\varepsilon_3 \sum_{jn} \left(\hat{b}_{jn} + \hat{b}_{jn}^\dagger\right)\left(\hat{b}'_{jn} + \hat{b}'_{jn}^\dagger\right).$$

The effects of the coupling to B' would be negligible in the limit of $M' \gg M$, and Eq. (C11) would reduce to Eqs. (19) and (20). Eq. (C11) is somewhat analogous to the symmetry breaking that mixes the three types of neutrinos and is responsible for neutrino oscillations [50]. The main differences are that the oscillaton family would have to be bosons and their number would not be conserved.

This example of a generalized theory does not maintain the usual commutation relations between the vector potential and the electric field for $\gamma \neq 0$, and the theory may be nonlocal as a result. A modified form of the theory that avoids these difficulties will be described in a subsequent paper.

This model is obviously speculative in the absence of any evidence for the existence of oscillations $(\gamma \neq 0)$. But it does provide some motivation for the form of the Bogoliubov transformation of Eq. (19), and the theory could be tested using the photon scattering experiment described in the next appendix.

## APPENDIX D: PHOTON SCATTERING

As briefly discussed in the text, the interaction Hamiltonian of Eqs. (18) and (19) allows a photon scattering process in which an incident photon with frequency $\omega$ is absorbed by a two-level atom into a virtual state, followed by the emission of a scattered photon at a frequency $\omega'$ along with the creation of a pair of oscillatons. In this appendix, the rate of subharmonic scattering of that kind is calculated using perturbation theory and compared with that for the usual elastic scattering of a photon. The effects of a nonzero oscillaton mass will also be considered.

In second-order perturbation theory, the rate $\Gamma$ of transitions from an initial state $|0\rangle$ to a final state $|f\rangle$ is given by [1]

$$\Gamma = \frac{2\pi}{\hbar} | \sum_l \frac{\langle f|\hat{H}'|l\rangle\langle l|\hat{H}'|0\rangle}{E_0 - E_l + i\eta\hbar}|^2 \ \delta(E_f - E_0), \quad \text{(D1)}$$

where the sum is over all intermediate states $|l\rangle$. $E_0$ is the energy of the initial state while $E_l$ and $E_f$ are the energies of the intermediate and final states.

The initial and final states of interest are illustrated in Fig. 8. The initial state $|0\rangle$ in Fig. 8(a) contains a single photon of frequency $\omega$ and wave vector **k**, along with an atom in its ground state. It also contains a single oscillaton in both of the modes corresponding to $\omega$ and $\omega'$, which is assumed to be the case for all of the modes in the initial state of the third-quantized theory. The oscillaton in the $\omega$ mode is in its first excited state, while the $\omega'$ mode is in its ground state, which corresponds to no photons in that mode.

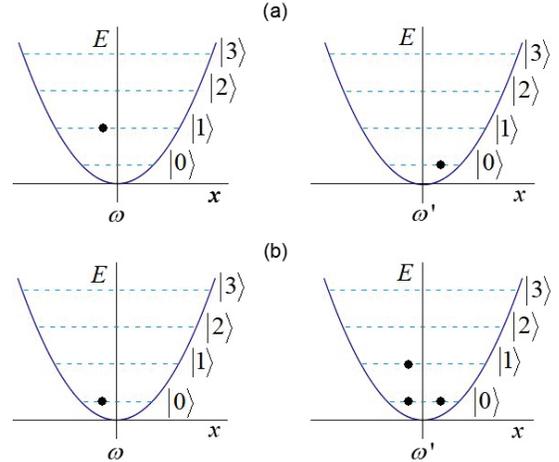

FIG. 8. Initial (a) and final (b) states of interest in the photon scattering experiment of Fig. 7. Each of the drawings represents the harmonic oscillators that corresponds to modes of the electromagnetic field with angular frequency $\omega$ or $\omega'$. The black dots show the energy levels occupied by the oscillatons in each mode.

The final state contains a scattered photon at frequency $\omega'$ and wave vector **k'**, along with an additional pair of oscillatons in that mode and the atom back in its ground state. This is illustrated in Fig. 8(b). Since there was one oscillaton initially in the $\omega'$ mode, there will now be a total of three, two in their ground state and one in its first excited state, which corresponds to the scattered photon. The oscillaton in mode $\omega$ will now be in its ground state, corresponding to no photons in that mode.

The incident photon will be assumed to have a linear polarization $\boldsymbol{\varepsilon}$ that is perpendicular to the plane containing the initial and final directions of photon propagation. It will also be assumed that the detuning $\Delta = (\hbar\omega - E_A)$ is much less than $\hbar\omega$, where $E_A$ is the energy of the first excited state of the atom relative to the ground state.

There are a number of possible intermediate states, but by far the largest contribution for $\Delta \ll \hbar\omega$ comes from a near-resonance interaction in which the intermediate state contains only the excited atomic state with all of the oscillatons in their lowest energy level (no photons). In that case, the energy difference that appears in the denominator of Eq. (D1) is given by $E_0 - E_l = \Delta$, and the corresponding matrix element is given in the dipole approximation [1] by

$$\langle l|\hat{H}'|0\rangle = i\omega q \left(\frac{2\pi\hbar}{\omega L^3}\right)^{\frac{1}{2}} \mathbf{d}\cdot\boldsymbol{\varepsilon}, \qquad (D2)$$

where $\mathbf{d} = \langle l|\mathbf{x}|0\rangle$ is the dipole moment of the atomic transition. Eq. (D2) is based on the $\cos\gamma \approx 1$ terms in Eq. (19) and it corresponds to the usual matrix element in quantum optics.

From Eqs. (7), (8), (18) and (19), the matrix element corresponding to the final transition is given by

$$\langle f|\hat{H}'|l\rangle = -i2\sqrt{2}\omega q\gamma \left(\frac{2\pi\hbar}{\omega' L^3}\right)^{\frac{1}{2}} \mathbf{d}\cdot\boldsymbol{\varepsilon}' \qquad (D3)$$

for $\gamma \ll 1$. The factor of $\omega$ comes from the matrix element of the current operator in the dipole approximation, which is a property of the atomic transition, while the factor of $\omega'$ comes from the definition of the vector potential operator. There can also be transitions to more complicated final states that are beyond the intended scope of this paper.

We will consider only those photons that are scattered into a small solid angle $d\Omega_S$ at a right angle to the incident beam, as illustrated in Fig. 7. With this geometry $\mathbf{d}\cdot\boldsymbol{\varepsilon} = \mathbf{d}\cdot\boldsymbol{\varepsilon}' = d$, where $d$ is the magnitude of the dipole moment. Inserting the matrix elements into Eq. (D1) and converting the sum over $\mathbf{k}'$ into an integral over $\omega'$ in the usual way [1] gives

$$\Gamma = 16\pi\alpha^2\gamma^2 \left(\frac{\hbar\omega}{\Delta}\right)^2 \left(\frac{d^4}{L^3\lambda}\right)\omega' d\Omega_S. \qquad (D4)$$

Here $\alpha = q^2/\hbar c$ is the fine structure constant and $\lambda$ is the wavelength of the incident photon. Eq. (D4) could be converted to a cross section, but that is not necessary for comparing the two scattering rates.

Including the zero-point energies of the oscillatons, the energies of the initial and final states are given by

$$\begin{aligned}E_0 &= \frac{3}{2}\hbar\omega + \frac{1}{2}\hbar\omega' \\ E_f &= \frac{1}{2}\hbar\omega + \frac{5}{2}\hbar\omega'.\end{aligned} \qquad (D5)$$

Setting $E_f = E_0$ and solving for $\omega'$ gives

$$\omega' = \frac{1}{2}\omega, \qquad (D6)$$

as discussed in the text.

The conventional elastic scattering rate can be calculated in the same way using second-order perturbation theory. The only significant differences in the calculation are that $\omega' = \omega$ and the factor of $2\sqrt{2}\gamma$ does not appear in the matrix element for the final transition. Taking the ratio of these two scattering rates gives $R = 4\gamma^2$ in agreement with Eq. (21) in the text.

Up to this point, it has been tacitly assumed that the oscillaton mass $m$ is either zero or negligibly small, since a photon has zero mass. If we include the possibility that $m \neq 0$, then energy conservation gives $\omega' = \omega/2 - mc^2/\hbar$. As a result, the photon scattering experiment of Fig. 7 could determine the mass of the oscillaton as well as the value of the mixing angle $\gamma$, provided that any scattering of that kind is observed.

**References**

1. G. Baym, *Lectures on Quantum Mechanics* (Benjamin, Reading, 1969).
2. L. Schiff, *Quantum Mechanics* (McGraw-Hill, New York, 1955).
3. J.J. Sakurai and J. Napolitano, *Modern Quantum Mechanics*, 2nd ed. (Cambridge University Press, Cambridge, 2017).
4. B.R. Desai, *Quantum Mechanics with Basic Field Theory* (Cambridge U. Press, Cambridge, 2010).
5. R.H. Landau, *Quantum Mechanics II*, 2nd ed. (Wiley, New York, 1996).

6. P. Roman, *Advanced Quantum Theory* (Addison-Wesley, Reading, 1965).
7. E. Koch, in *Emergent Phenomena in Correlated Matter*, E. Pavarini, E. Koch, and U Schollwock, eds. (Institute for Advanced Simulation, Julich, 2013.)
8. G.S. Agarwal, *Quantum Optics* (Cambridge U. Press, Cambridge, 2013), p. 7.
9. S.M. Barnett and P.M. Radmore, *Methods in Theoretical Quantum Optics* (Oxford U. Press, Oxford, 1997), p. 64.
10. U. Leonhardt, *Measuring the Quantum State of Light* (Cambridge U. Press, Cambridge, 1997), p. 20.
11. W. Schleich, *Quantum Optics in Phase Space* (Wiley, Berlin, 2001), p. 308.
12. S.U. Shringarpure and J.D. Franson, Phys. Rev. A **102**, 023719 (2020).
13. W.A. Harrison, *Solid State Theory* (McGraw-Hill, New York, 1970).
14. S.M. Girvin and K. Yang, *Modern Condensed Matter Theory* (Cambridge U. Press, Cambridge, 2019).
15. M. El-Batanouny, *Advanced Quantum Condensed Matter Physics: One-Body, Many-Body, and Topological Perspectives* (Cambridge U. Press, Cambridge, 2020).
16. J.R. Schrieffer, *Theory of Superconductivity* (W.A. Benjamin, Reading, 1964).
17. R.D. Parks, ed., *Superconductivity*, Vols. I and II (Marcel Dekker, New York, 1969).
18. C. Cohen-Tannoudji, J. Dupont-Roc, and G. Grynberg, *Photons and Atoms: Introduction to Quantum Electrodynamics* (Wiley, New York, 1989).
19. W. Heitler, *The Quantum Theory of Radiation*, 3rd ed. (Dover, Mineola, 1984).
20. M.O. Scully and M. S. Zubairy, *Quantum Optics* (Cambridge U. Press, Cambridge, 1997).
21. Y. Nambu, Prog. Theor. Phys. **4**, 331 (1949).
22. M. McGuigan, Phys. Rev. D **38**, 3031 (1988).
23. A. Strominger, Phil. Trans. R. Soc. Lond. A **329**, 395 (1989).
24. Y. Xiang and L. Liu, Chin. Phys. Lett. **8**, 52 (1991).
25. Y. Peleg, Class. Quantum Grav. **8**, 827 (1991).
26. T. Prosen, New J. Phys. **10**, 043026 (2008).
27. L.A. Urena-Lopez, Class. Quantum Grav. **19**, 2617 (2002).
28. M.P. Hertzberg, Phys. Rev. D **82**, 045022 (2010).
29. G. Fodor, P. Forgacs, and M. Mezei, Phys. Rev. D **81**, 064029 (2010).
30. M.J. Collett, R. Loudon, and C.W. Gardiner, J. Mod. Optics **34**, 881 (1987).
31. C.M. Caves, Phys. Rev. D **23**, 1693 (1981).
32. D.F. Walls, Nature **306**, 141 (1983).
33. B.T. Kirby and J.D. Franson, Phys. Rev. A **87**, 053822 (2013).
34. S. Stenholm, Physica Scripta **T12**, 56 (1986).
35. A.A. Clerk, M.H. Devoret, S.M. Girvin, F. Marquardt, and R.J. Schoelkopf, Rev. Mod. Phys. **82**, 1155 (2010).
36. C.M. Caves, J. Combes, Z. Jiang, and S. Pandey, Phys. Rev. A **86**, 063802 (2012).
37. J.D. Franson and R.A. Brewster, Phys. Lett. A **382**, 887 (2018). Cross-terms such as $\psi^*(x_1)\psi(x_2)$ in Eq. (24) of this reference (in the Schrodinger picture) are equivalent to the Heisenberg operator $\hat{C}$ in this paper.
38. S. Weinberg, Phys. Rev. Lett. **13**, 495 (1964).
39. T.G. Rizzo, Phys. Rev. D **34**, 3519 (1986).
40. M. Suzuki, Phys. Rev. Lett. **56**, 1339 (1986).
41. W. Schleich, op. cit., p. 307.
42. J.D. Franson and R.A. Brewster, arXiv.1811.06517 (2018).
43. N.N. Bogoliubov, Il Nuovo Cimento **7**, 794 (1958).
44. J.G. Valatin, Il Nuovo Cimento **7**, 843 (1958).
45. C. Emary and R.F. Bishop, J. Math. Phys. **43**, 3916 (2002).
46. A.I. Lvovsky, W. Wasilewski, and K. Banaszek, J. Mod. Opt. **54**, 721 (2007).
47. W.G. Unruh, Phys. Rev. D **14**, 870 (1976).
48. N. Cabibbo, Phys. Rev. Lett. **10**, 531 (1963).
49. S. Weinberg, Phys. Rev. Lett. **19**, 1264 (1967).
50. D. Griffiths, *Introduction to Elementary Particles* (Wiley, Weinheim, 2008).
51. L. Gurung et al., Phys. Rev Lett. **125**, 073002 (2020).
52. M. Rini, Physics **13**, s99, doi.org/10.1103/Physics.13.s99 (2020).